# Structural and magnetic characterization

# of self-assembled iron oxide nanoparticle arrays

M. J. Benitez[1,2], D. Mishra[1], P. Szary[1], G.A. Badini Confalonieri[1],

M. Feyen[2], A. H. Lu[2], L. Agudo[3], G. Eggeler[3], O. Petracic[1] and H. Zabel[1]

[1]Institut für Experimentalphysik/Festkörperphysik, Ruhr-Universität Bochum, D-44780 Bochum, Germany

[2]Max-Planck Institut für Kohlenforschung, D-45470 Mülheim an der Ruhr, Germany

[3]Institut für Werkstoffe, Ruhr-Universität Bochum, D-44780 Bochum, Germany

E-mail: Maria.BenitezRomero@ruhr-uni-bochum.de, Durga.Mishra@ruhr-uni-bochum.de, and Oleg.Petracic@ruhr-uni-bochum.de

**Abstract:** We report about a combined structural and magnetometric characterization of self-assembled magnetic nanoparticle arrays. Monodisperse iron oxide nanoparticles with a diameter of 20 nm were synthesized by thermal decomposition. The nanoparticle suspension was spin-coated on Si substrates to achieve self-organized arrays of particles and subsequently annealed at various conditions. The samples were characterized by x-ray diffraction, bright and dark field high resolution transmission electron microscopy (HRTEM). The structural analysis is compared to the magnetic behavior investigated by superconducting interference device (SQUID) magnetometry. We can identify either multi-phase $Fe_xO/\gamma$-$Fe_2O_3$ or multi-phase $Fe_xO/Fe_3O_4$ nanoparticles. The $Fe_xO/\gamma$-$Fe_2O_3$ system shows a pronounced exchange bias effect which explains the peculiar magnetization data obtained for this system.

## I. Introduction

Magnetic nanoparticles (NPs) are in the focus of much interest because of their potential use as building blocks for future high-density data storage systems [1-4], spintronic devices [5, 6], and for photonic [7], bio-medical [8-12] or refrigeration applications [13]. Rapid developments in the chemical synthesis of magnetic NPs offer the possibility to exploit novel magnetic, optical and electrical properties that emerge when reducing the size of the particles. NPs can be synthesized with controlled size, shape and surface coating. In particular iron oxide NPs are currently intensely discussed for medical applications due to their bio-compatibility and for spintronic devices due to their half-metallic properties [8, 12, 14-17]. Applications, however, require the precise identification of the iron phase, which is often neglected. Therefore, we employed a combined structural and magnetometric characterization of iron oxide NP arrays at three different annealing conditions, i.e. moderate drying at 80°C in air (as-prepared state), annealing at 170°C in air and annealing at 230°C in vacuum. This yields either multi-phase $Fe_xO/\gamma$-$Fe_2O_3$ or multi-phase $Fe_xO/Fe_3O_4$ NPs. Annealing at 170°C in air seems to produce single phase $\gamma$-$Fe_2O_3$ considering only structural information. However, using a more detailed investigation using remanent magnetization curves a minor $Fe_xO$ phase is detected here as well.

'Wüstite' ($Fe_xO$) is a nonstoichiometric phase with a known stability range from $x = 0.83$ to 0.96. At room temperature wüstite is paramagnetic and crystallizes in a rock salt structure, which is a close-packed fcc $O^{2-}$ lattice with $Fe^{2+}$ ions occupying the B interstitial sites. Below the Néel temperature $T_N = 198$ K, wüstite orders antiferromagnetically (AF) [18, 19]. The magnetic moments order in two-dimensional Ising-like ferromagnetic (FM) sheets parallel to the (111) planes with the moment directions pointing alternately up and down in adjacent sheets [18]. The paramagnetic to AF transition is accompanied by a slight elongation along the [111] direction where the crystal becomes rhombohedral. The magnetic ordering can be understood in terms of a predominantly AF coupling between next-nearest neighbor ions in the [100] lattice directions.

'Magnetite' ($Fe_3O_4$) is a ferrimagnetic (FiM) compound below 858 K. It crystallizes in a cubic 'inverse spinel' structure [19-21] with the tetrahedrally coordinated A sites occupied by $Fe^{3+}$, and the octahedrally coordinated B sites equally occupied by iron atoms with formal +3 and +2 charges. The $Fe^{2+}$ and $Fe^{3+}$ ions in the octahedral sites are aligned FM by the double exchange interaction, whereas the $Fe^{3+}$ ions on the tetrahedral sites are coupled to the $Fe^{3+}$ in the octahedral sites by an AF superexchange interaction. In contrast to other ferrites, magnetite is a relatively good conductor at room temperature. The conductivity is associated with the mixed valency, which gives rise to FM exchange interactions. Oxides containing a single valency, $Fe^{2+}$ or $Fe^{3+}$, are magnetic insulators. The conductivity of bulk magnetite at room temperature is approximately 200 $(\Omega cm)^{-1}$ [22]. When cooling below the characteristic Verwey transition at $T_V$ = 120 K its conductivity drops by two orders of magnitude accompanied by a slight crystallographic distortion [20, 21, 23].

'Maghemite' ($\gamma$-$Fe_2O_3$) is a FiM material below 948 K [19]. It crystallizes in the inverse spinel structure similar to magnetite. In contrast to magnetite, eight $Fe^{3+}$ ions are located in tetrahedral sites (A-sites) and sixteen $Fe^{3+}$ occupy the octahedral sites (B-sites). The FiM property in maghemite is the result of $Fe^{3+}$ in the B sites of the spinel structure. The saturation magnetization of maghemite is 380 kA/m which is smaller compared to 480 kA/m of magnetite.

Apart from the above mentioned oxide phases there are also the phases $\alpha$-$Fe_2O_3$, $\beta$-$Fe_2O_3$ and $\varepsilon$-$Fe_2O_3$. Among those, 'hematite', ($\alpha$-$Fe_2O_3$) is the most abundant in nature. The other two phases are metastable and are formed by reduction of $\alpha$-$Fe_2O_3$ at high temperatures. Hematite is AF below 948 K [19]. It also shows a transition from weak ferromagnetism to AF at 260 K, known as Morin transition. It crystallizes in a rhombohedral structure at room temperature and from x-ray diffraction measurements one can easily distinguish it from magnetite, maghemite and wüstite. In our studies hematite was not identified for all annealing temperatures.

## II. EXPERIMENTAL DETAILS

Spherical monodisperse iron oxide NPs with oleic acid coating were synthesized by thermal decomposition of iron oleate in trioctylamine in presence of oleic acid following the procedure described by Park *et al* [24]. In a typical synthesis, the metal-oleate complex was prepared by reacting 3 g of iron chloride ($FeCl_3*6H_2O$) and 10.13 g of sodium oleate with 22.21 ml ethanol, 16.62 ml water and 38.88 ml *n*-hexane. The mixture was heated to 70°C for 4 h. In the next step, 6 g of iron oleate and 0.96 ml of oleic acid were added to 43.53 ml of trioctylamine and stirred in a three-neck round-bottom flask. The mixture was heated to 320°C with a heating rate of 3.3°C/min under vigorous stirring. Once the temperature was reached, the reaction mixture was kept at that temperature for 30 min. The initial reddish-brown color of the reaction solution turned brownish-black. The resultant solution was then cooled to room temperature. The thus prepared NPs were separated by centrifugation and washed with ethanol. This procedure was repeated five times. After washing, the resultant NPs were separated by centrifugation and dissolved in toluene for long-term storage. Hereby the NPs remain coated with the oleic acid shell.

Next, the iron oxide NPs were spin-coated on Si(100) substrates with a native oxide. The substrates were ultrasonically cleaned in acetone for 15 min, rinsed with isopropanol and dried with pure $N_2$ stream. In a typical procedure, 0.2 ml of iron oxide NP solution was spun at 3000 rpm for 30 s and dried on a hot plate at 80 °C for 20 min.

Structural characterization of the NP monolayer films was carried out by electron microscopy and x-ray diffraction. Transmission electron microscopy (TEM) images were obtained with a HF 2000 microscope (Hitachi) equipped with a cold field emission gun and with a QUANTA 200 FEG scanning electron microscope (SEM), respectively. High resolution TEM (HRTEM) and selected area electron diffraction (SAED), conventional dark field and bright field images were taken with an Analytical 200kV FEG-TEM TECNAI F20 S-Twin instrument. X-ray Bragg scans were carried out at beamline BL09 at the DELTA synchrotron facility in Dortmund, Germany. The photon energy was selected to be 11 keV ($\lambda$= 1.127 Å) using a Si (311) double crystal monochromator. The angle of incidence to the film was kept fixed at 0.2°, which is just above the critical angle for total reflection. The beam size at the sample position was 0.2mm x 2mm. The $2\theta$ angle of the detector was scanned from 10° to 60° with a step size of 0.05°. Magnetometry measurements of the samples were performed using a Quantum Design MPMS5 superconducting quantum interference device (SQUID) magnetometer in applied magnetic fields up to 50 kOe.

## III. RESULTS AND DISCUSSION

### A. Structural characterization

Figure 1 (a, b) shows transmission electron microscopy (TEM) images of the as-prepared NPs. The images reveal the formation of highly monodisperse particles with an average size of 20 nm and relatively narrow size distribution of 7%. Figure 1 (c, d) shows SEM images of the self-assembled NPs on a Si substrate with approximately ten monolayers (a) and one monolayer (b), showing densely packed hexagonal arrays. The number of layers of NPs was controlled by changing the concentration of the solution.

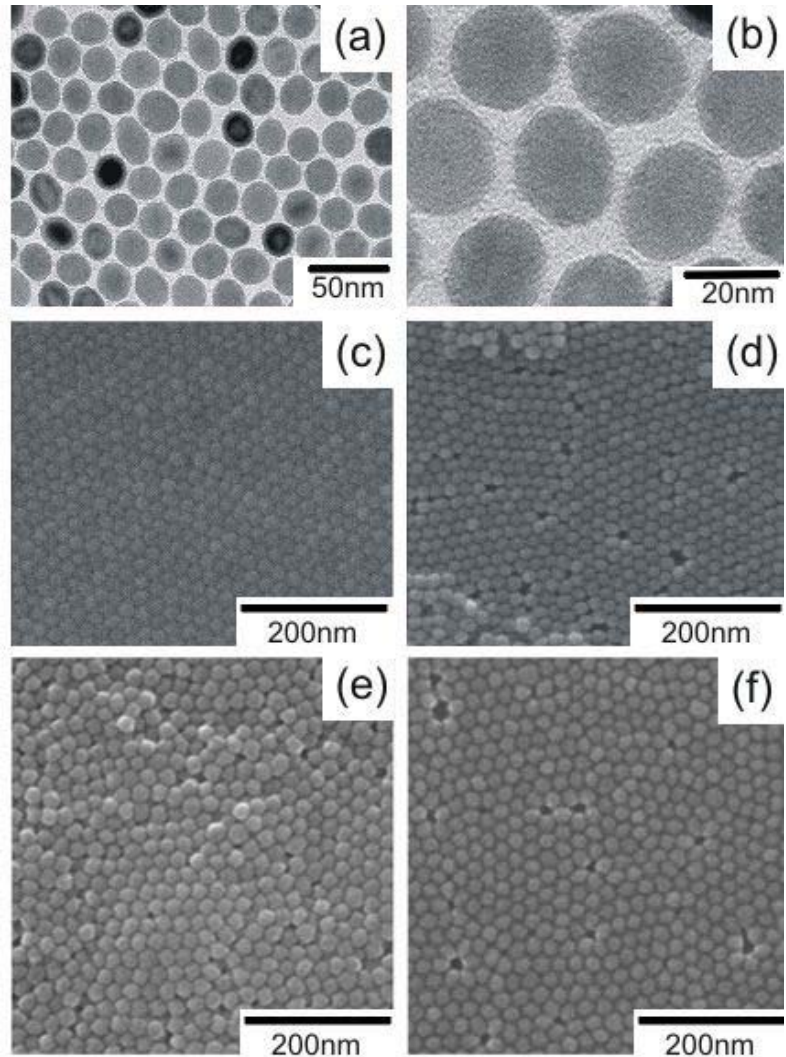

**FIG. 1**: (a, b) TEM images of as-prepared iron oxide NPs with a diameter of 20 nm. SEM images of (c) a multilayer and (d) a monolayer of NPs on Si substrates dried at 80°C and (e) a multilayer and (f) a monolayer of NPs annealed at 230°C in vacuum, respectively.

Figure 2(a) shows a powder x-ray diffraction pattern of the samples dried at 80 °C. Analysis of the diffraction pattern gives an indication of the presence of a mixture of iron oxide phases. The broad peak at 25.7° hints toward the presence of two crystalline phases, wüstite and spinel. The main diffraction peaks (111), (200) and (220) of $Fe_xO$ are clearly identified. However, one observes diffraction peaks at (220) and (511), which correspond to the spinel $F_3O_4$ or $\gamma$-$Fe_2O_3$.

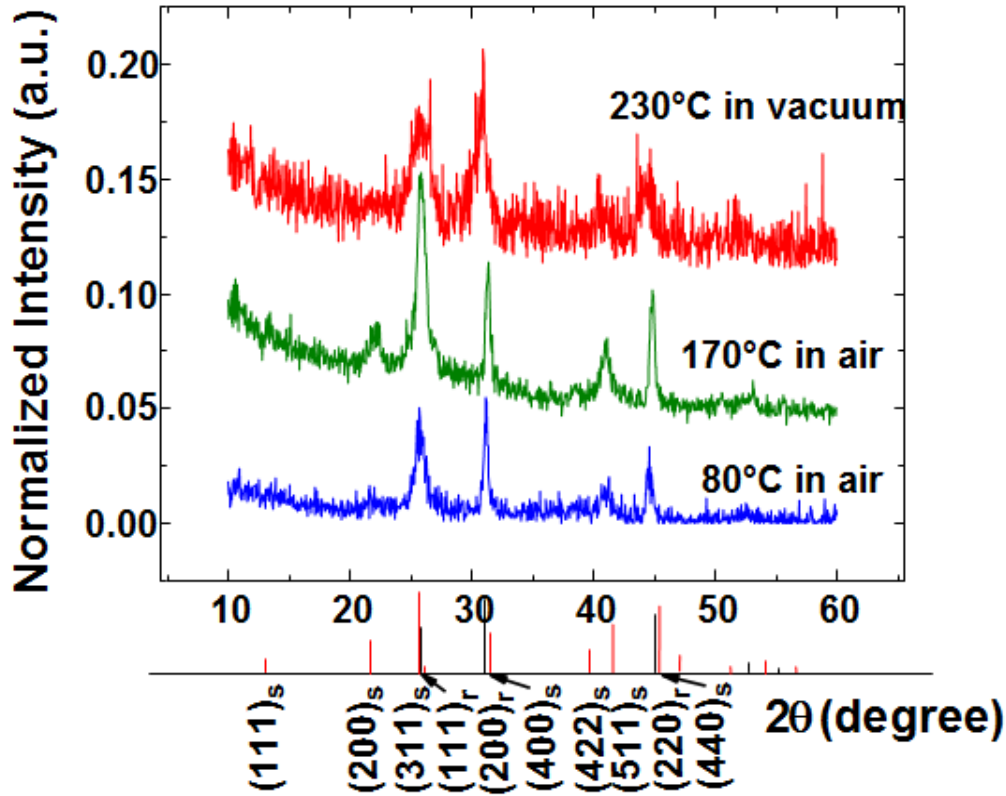

**FIG. 2.** XRD patterns of NPs on Si substrates, (a) dried in air at 80°C, (b) annealed in air at 170°C and (c) in vacuum at 230°C. On the horizontal bar below the position and intensity of powder Bragg peaks are shown according to literature values. The Miller indices are labeled by 's' for the spinel structure and 'r' for the rock salt (wüstite) structure.

The mixture of iron phases in one NP is attributed to the surface oxidation of the $Fe_xO$ nanoparticles [25] resulting in the formation of a $F_3O_4$ layer around the nanoparticles [24], or due to an incomplete removal of the surfactants inside the NPs during the synthesis, leading to a lower intensity in the middle of the particle that inhibit the formation of one single iron oxide phase [26]. By annealing in oxidizing atmospheres, it has been reported that wüstite nanocrystals can be transformed into magnetite or maghemite nanocrystals [24, 25, 27]. One should note that we find no significant difference between samples dried at 80°C and samples without heat treatment. Therefore, the 80°C annealed systems are identical to the as-prepared systems with respect to crystallographic structure. However, we find a better adhesion of the NPs to the substrate for the system dried at 80°C.

For the second heat treatment, NP films were annealed on a hot plate at 170 °C for 20 min (with an intermediate plateau at 80 °C for 20 min.). SEM images from the resulting films (not shown here) indicate that the particles did not coalesce during annealing at 170 °C. In Figure 2(b) a powder diffraction pattern of the samples annealed at 170 °C is shown. The $Fe_xO$ phase undergoes a chemical and structural conversion into $F_3O_4$ or $\gamma$-$Fe_2O_3$ evidenced by the shift of the diffraction peaks to high angles, increase of intensities of (200), (311) and (511) and appearance of a (422) peak. These results favor the presence of a spinel structure as single or major phase in the NPs.

The third heat treatment consisted of annealing the as-prepared NP films in high vacuum ($2\times10^{-7}$ mbar) at 230 °C for 20 minutes. The heating and cooling was performed inside a vacuum chamber. SEM images show that the particles still retain the close-packed structure without any neck formation [28]. Fig 2 (c) shows the x-ray diffraction pattern of a multilayer NP film. It is similar to the pattern of the as-prepared system, which indicates a mixed phase. The (311), (511), and (440) peak of $F_3O_4$ or $\gamma$-$Fe_2O_3$ are clearly visible. The (111), (200) and (220) peaks of the fcc wüstite phase can also be identified.

For TEM analysis samples were prepared on $SiO_2$ coated copper grids and heat treated as before at 80°C in air, 170°C in air or 230°C in vacuum for 20 minutes. We used $SiO_2$ coated grids to maintain similar conditions for the self-assembly of NPs as in the case of NP films prepared on Si substrates with a natural $SiO_2$ layer.

Fig 3(a) shows the bright field TEM image of the NPs, (b) the corresponding selected area electron diffraction (SAED) pattern, (c-d) HRTEM images and (e-h) the conventional dark field images for the samples annealed at 80°C in air.

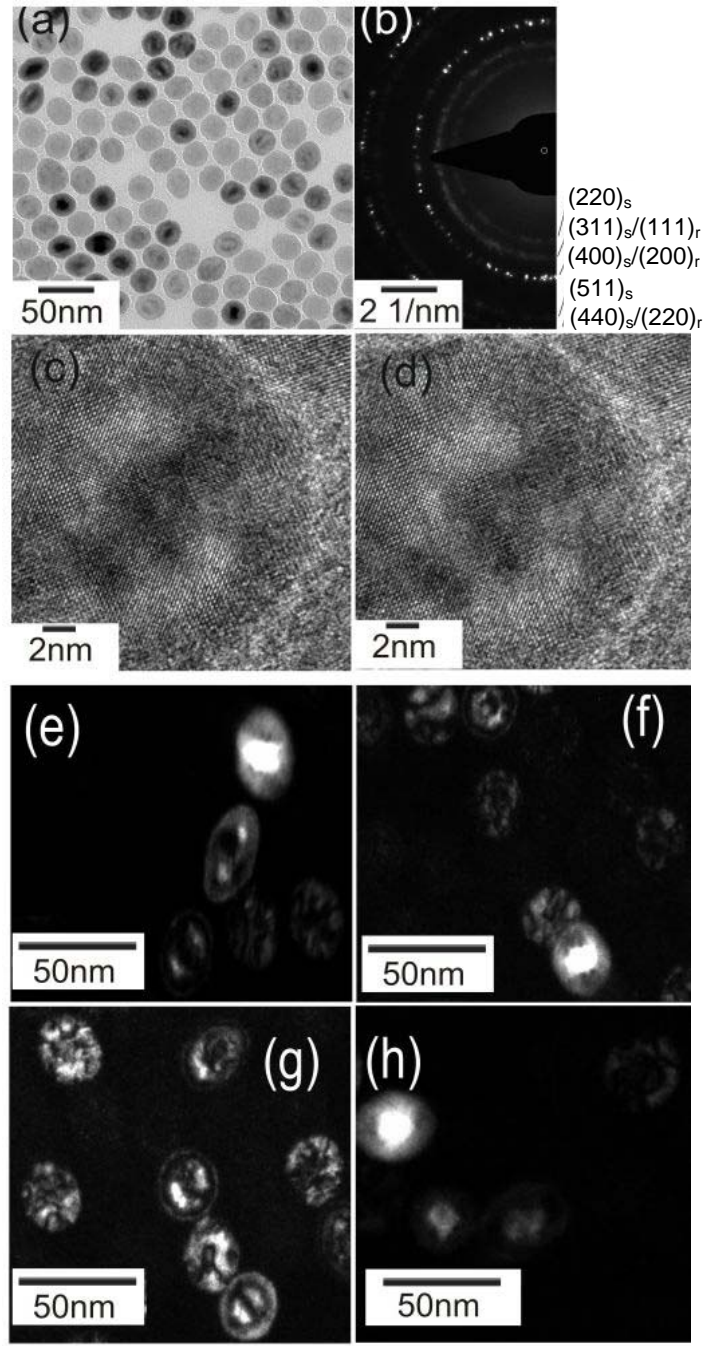

**FIG 3**. (a) Bright field TEM image of NPs annealed at 80°C in air (scale bar corresponds to 50nm). (b) Corresponding SAED pattern (the subscripts s and r in the indexing stands for inverse spinel magnetite or maghemite and rock salt wüstite respectively). (c) and (d) show HRTEM images at two different focuses showing the polycrystalline nature of each particle (scale bar corresponds to 2nm). (e), (f), (g) and (h) are the dark field TEM images taken at different ring positions showing a multi-phase structure of spinel phase and wüstite (scale bar corresponds to 50nm).

The bright field image shows mostly spherical particles with relatively small size distribution of $\approx 7\%$. The SAED pattern reveals the presence of an inverse spinel structure (maghemite or magnetite) and the rock salt structure of wüstite, confirming our conclusions from the x-ray diffraction pattern. After carefully indexing the textured SAED rings we notice that the first intense broad ring has a contribution from the (311) peak of the spinel structure (magnetite or maghemite) and from the (111) peak of the rock salt structure (wüstite). The other two intense rings could be assigned to the (200) and (220) planes of wüstite. The (511) and (220) planes of the spinel structure could also be observed in this case. The example HRTEM images (c) and (d) demonstrate that the NP ensemble consists of a mixture of single crystals with several coherent phases or polycrystalline particles.

We analyzed dark field images by choosing particular diffraction spots from the diffraction ring. Although the diffraction rings are very close for wüstite and inverse spinel structure, a clear multi-phase wüstite-spinel structure can be observed (Fig 3 (e), (f), (g), (h)). The diffraction rings $(440)_s/(220)_r$, $(511)_s$ and $(422)_s$ contribute to the dark field images taken at different ring positions. The aperture does not allow imaging of a single diffraction ring. Nevertheless one can observe one particle with a multi-phase structure. Considering another spot on the same ring (h) a different particle phase is illuminated. In panel (f), which belongs to the same ring, we do not encounter a clear multi-phase structure. We conclude that, in general, particles are composed of small crystallites (grains) of different phases (wüstite or inverse spinel) with variable volume ratio from particle to particle.

Fig 4(a) shows the bright field TEM image of NPs annealed at 170°C in air. The HRTEM image in (c) reveals a single-crystalline NP. One should note that the inhomogeneous contrast is probably due to focus artifacts introduced by the shape of the NP. Image (d) shows, however, a polycrystalline particle. The corresponding SAED pattern is presented in panel (b). The diffraction rings correspond to the inverse spinel maghemite or magnetite phase indicating basically single-phase iron oxide supporting the x-ray diffraction study.

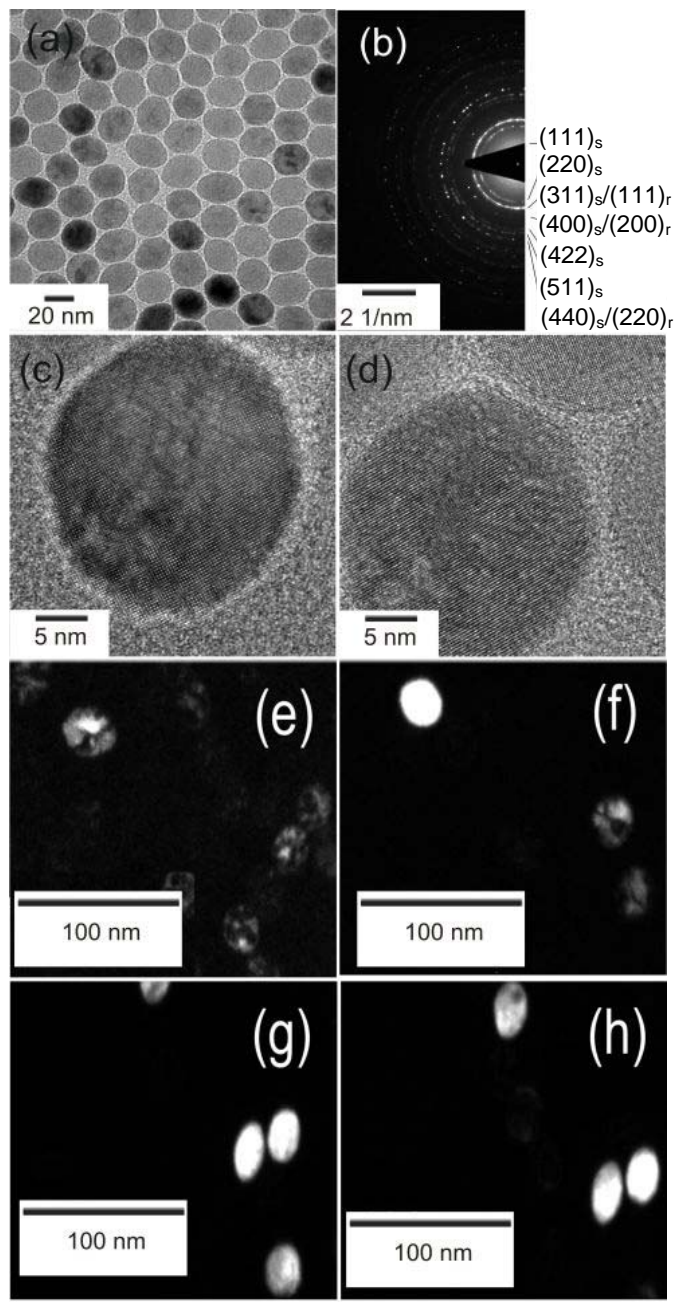

**FIG 4**. (a) Bright field TEM image of NPs annealed at 170°C in air (scale bar corresponds to 20nm). (b) Corresponding SAED pattern. (c) and (d) show HRTEM images of two different particles (scale bar corresponds to 5nm). (e), (f), (g) and (h) are the dark field TEM images taken at different ring positions showing a multi-phase structure of spinel phase and wüstite (scale bar corresponds to 100nm).

The dark field image in panel (e) taken from $(220)_s$, $(311)_s$ and partly from $(200)_r$/ $(400)_s$ rings again show a mixed-phase structure. The dark field image in panel (f) shows a similar trend.

Images in panel (g) and (h) are taken from $(422)_s$, $(511)_s$ and $(220)_r/ (440)_s$. They show predominantly NPs of single phase. Hence, no clear conclusion about the single- crystalline or single-phase nature can be drawn from structural data alone in the case of NPs annealed at 170°C in air. The structural analysis indicates only a mixture of single-crystalline or polycrystalline NPs.

Fig 5(a) shows the bright field TEM image of NPs annealed at 230°C in vacuum. The corresponding SAED pattern is depicted in panel (b). The diffraction rings are not clearly defined, because of a larger illuminated area. One should note that the SAED patterns for the three annealing conditions were not taken from the same amount of particles. That means the intensities in each case cannot be compared directly. In the supplementary data file we show the TEM images and the corresponding illuminated areas which contributed to the SAED pattern in these three cases. The diffraction rings corresponding to the (200) and (220) planes of wüstite always exhibit discrete intensity spots indicative for a pronounced texture, whereas the diffraction rings corresponding to magnetite or maghemite are almost continuous as expected for a randomly distributed powder with negligible texture.

The HRTEM images in panels (c) and (d) show that also single-crystalline NPs can be encountered. However, the majority of NPs consists of a multi-phase structure. Fig 5 (e), (f), and (h) show dark field images taken from different positions on the ring $(311)_s/(111)_r$. Panel (g) shows the dark field image taken from ring $(400)_s/(200)_r$. Image (e) shows a core-shell particle. The core appears darker, because the spot of the diffraction ring is associated to the inverse spinel structure. Panels (f) and (g) depict NPs containing several small crystallites. Panel (h) shows one NP which is entirely illuminated evidencing a single-phase structure. Hence, the entire set of NPs consists of a mixture of various compositions with the majority of multi-phase particles with spinel and wüstite.

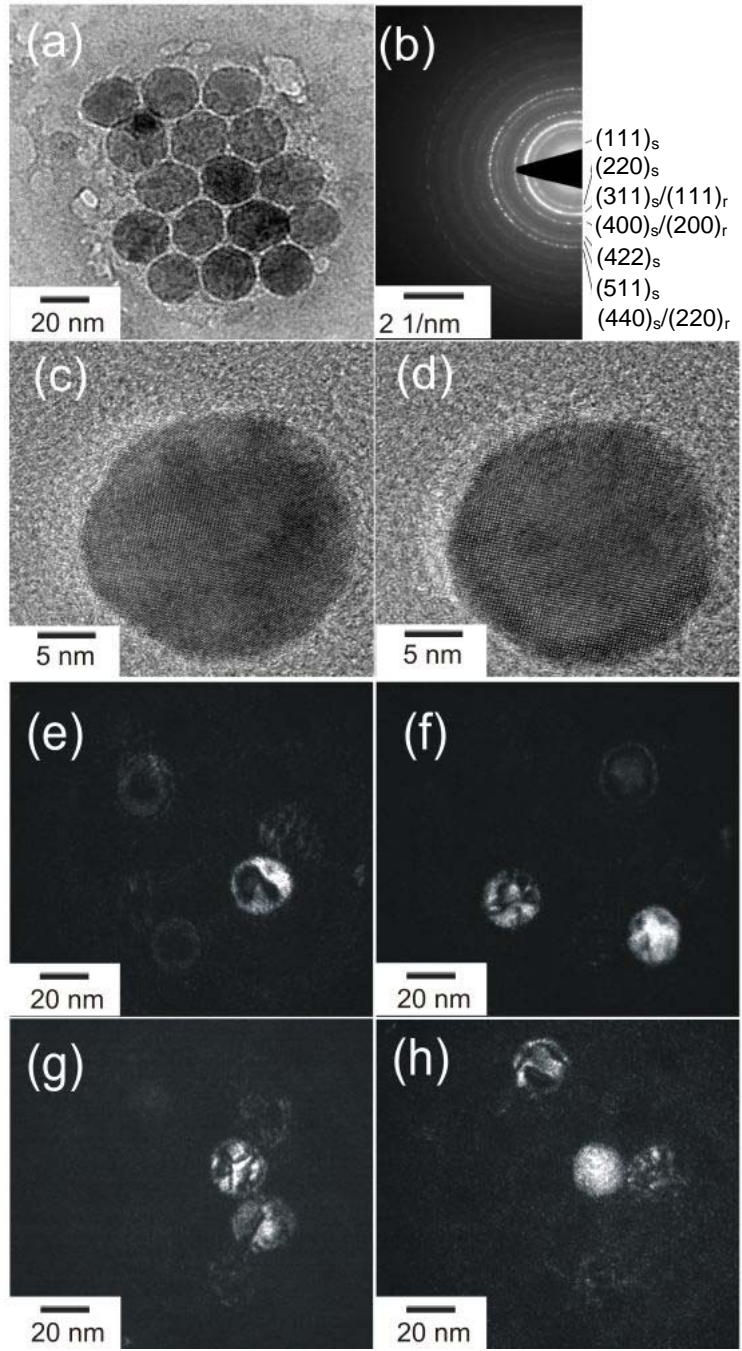

**FIG 5.** (a) Bright field TEM image of the NPs annealed at 230°C in vacuum (scale bar corresponds to 20nm). (b) Corresponding SAED pattern. (c) and (d) show HRTEM images displaying a polycrystalline mixed-phase structure (scale bar corresponds to 5nm). (e), (f), (g) and (h) show the dark field image of the particles (scale bar corresponds to 20nm).

## B. Magnetic characterization

Figure 6 (a) depicts *M vs. T* curves after zero field cooling (ZFC) and after field cooling (FC) measured at 50 Oe for a monolayer film of NPs dried at 80°C. The ZFC magnetization curve is obtained by first cooling the system in zero field from 330 K to 15 K. Next, the field is applied and subsequently the magnetization is recorded while increasing the temperature gradually. The FC magnetization curve is measured by decreasing the temperature in the same applied field.

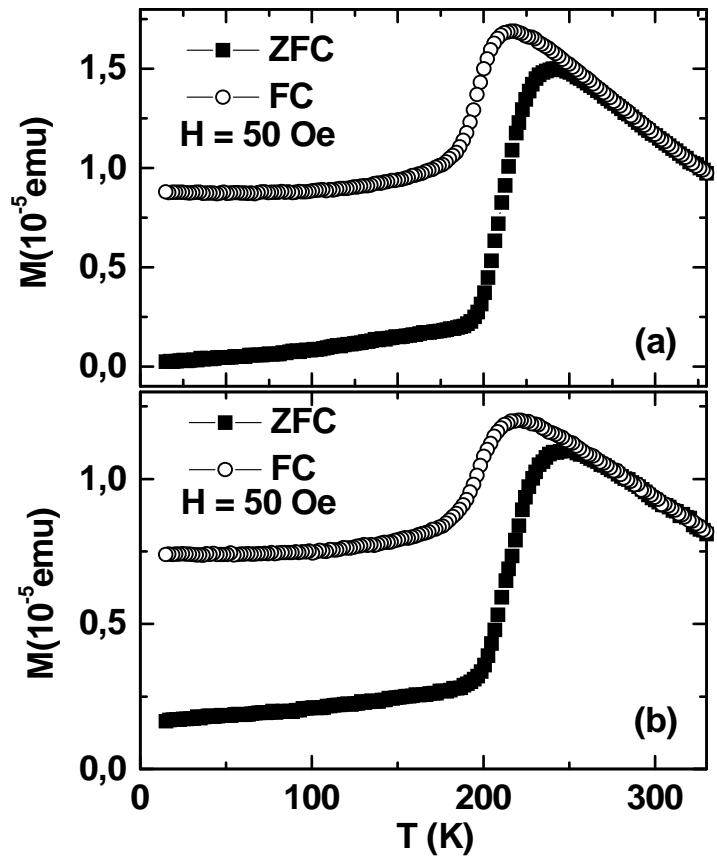

**FIG. 6.** *M* vs. *T* curves after ZFC and FC measured at 50 Oe for a monolayer of NPs after spin-coating of NP dispersions dissolved in toluene (a) and benzene (b). Both samples were dried at 80°C in air.

The ZFC and FC curves *seem* to correspond to those of a superparamagnetic (SPM) system [29-33], i.e. showing a gradual increase of the ZFC curve until a peak at the temperature $T_{p,ZFC}$ = 240 K and for higher temperatures a $1/T$ Curie-type decrease. Furthermore, the FC curve overlaps with the ZFC curve for high temperatures and splits from the ZFC curve below a bifurcation temperature $T_b \approx 260$ K slightly above $T_{p,ZFC}$. However, three features significantly deviate from the usual SPM picture. First, the ZFC curve shows a very atypical steep increase at a temperature $T_s \approx 200$ K. Second, the temperature of the peak in the FC curve, $T_{p,FC}$, is different from that of the ZFC curve. This behavior is atypical for a purely SPM systems. Third, the FC curve decreases below its peak temperature. Between $T_s$ and $T_{p,FC}$ the ZFC and FC curves are approximately parallel indicating a common phenomenon responsible for the shape of the curve.

The sudden increase in the magnetization at $T_s \approx 200$ K could be associated with either the freezing point of the solvent [34] or with a structural or magnetic transformation within the NPs. Or it could be due to two exchange coupled magnetic phases inside each NP.

In order to check whether this feature is caused by the freezing point of the solvent [34], further experiments were performed. Note that the NP films were dried at 80°C, which is below the boiling point of toluene (i.e. 110°C). Therefore, it is possible that the NPs were still surrounded by a residual toluene layer. A possible scenario could be that at low temperatures the liquid matrix surrounding the particles becomes frozen and thus mechanically fixing the NPs. By warming up the system above the melting point of the toluene (i.e. 180 K), the matrix would unfreeze. In the liquid matrix the particles would then be free to rotate.

Different samples were prepared under identical conditions, however, with benzene (melting point = 279 K) as solvent instead of toluene. Figure 6(b) shows ZFC and FC curves measured at 50 Oe for a monolayer film of NPs dissolved in benzene and dried at 80°C. The curves are basically identical to those presented in Fig. 6(a). Consequently the increase in magnetization close to 200 K is not due to the freezing and melting of a residual solvent matrix.

A better insight into the nature of the observed behavior can be drawn from $M$ vs. $T$ curves at various fields. In the case of a SPM or a superspin glass (SSG) system it is expected that the peak in the ZFC curve is strongly field dependent and shifts to lower temperatures for increasing fields (see Ref. 27-31 and references therein).

Figure 7 (a) shows the magnetization vs. temperature after ZFC and FC measured at relatively low fields of $H = 0.2$, 0.5, 1 and 2 kOe for a monolayer film of NPs dried at 80 °C. We indeed find that the peak in the ZFC curve shifts to lower temperatures as the field increases. However, the FC peak position remains constant for applied fields up to 2 kOe.

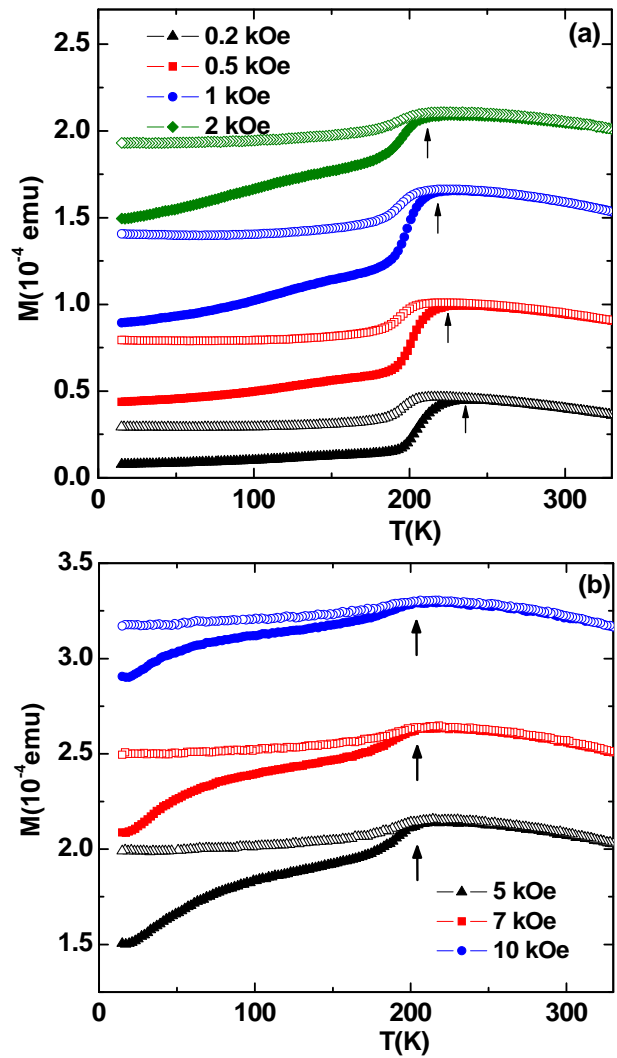

**FIG. 7.** *M* vs. *T* curves after ZFC (solid symbols) and FC (open symbols) measured at various applied fields of a monolayer of NPs dried at 80°C in air (a, b). The curves are shown with an offset of $3 \times 10^{-5}$ emu (measured at 0.5 kOe), $7 \times 10^{-5}$ emu (1 kOe) and $8 \times 10-5$ emu (2 kOe), respectively, for better clarity. Arrows in (a) mark the peak in the ZFC curve. Arrows in (b) mark the peak in the FC curve.

Figure 7 (b) shows ZFC and FC curves at higher fields $H$ = 5, 7 and 10 kOe. Again we observe a splitting between ZFC and FC curves just as for low fields. However, the ZFC and the FC peaks coincide in temperature and do not shift to lower temperatures with increasing field. This is contrary to the behavior expected for the blocking temperature ($T_B$) of SPM or the glass temperature ($T_g$) of SSG systems (see Ref. 27-31 and references therein). This can be explained by the presence of an AF.

In most AF systems the field dependence of the critical phase boundary is very small in the range of the usually accessible experimental field values. Furthermore, the ZFC and FC curves can have a very similar shape to that of SPM and SSG systems [35]. This matches well with the observation seen in Fig. 7 (b). Comparing the ZFC curves measured at $H > 2$ kOe, one finds virtually no change of the peak position at $T \approx 206$ K. Therefore, this temperature is very likely to be related to the Néel temperature of the AF wüstite phase, $T_N$ = 198 K (bulk value) [19]. Note that the structural characterization of the sample dried at 80°C using x-rays indicates the presence of a mixture of iron oxide phases, wüstite and spinel ($Fe_3O_4$ or $\gamma$-$Fe_2O_3$). The Néel temperature of the $Fe_xO$ phase might be increased compared to the bulk value possibly by the exchange interaction to the FiM phase inside the NPs.

We argue that the particular shape of the ZFC and FC curves is caused by a superposition of regular SPM blocking and exchange bias (EB) between AF and ferrimagnetic phases inside each NP [36, 37, 38].

From this finding we confirm that the NPs consist of $Fe_xO$ and spinel ($Fe_3O_4$ or $\gamma$-$Fe_2O_3$) phases. We also note that in the magnetization curves near 120 K no features are present that would mark the Verwey transition expected for $Fe_3O_4$ [20, 21, 23]. Evidence of the Verwey transition in nanocrystalline magnetite thin films [39] and in highly nonstoichiometric nanometric powders [40] were found using magnetic susceptibility measurements. Studies on magnetite NPs suggest that the Verwey transition shifts to lower temperatures due to the finite-size effect or vanishes below a critical size [17, 41]. Therefore, these samples either do not contain $Fe_3O_4$ or only a small fraction not showing the Verwey transition. Consequently our NPs are composed of only wüstite-maghemite.

The EB effect is confirmed by measurements of magnetization hysteresis loops. Fig. 8 shows $M$ vs. $H$ curves recorded at 15 K after ZFC and FC. The FC curves were measured by cooling the system from 330 K to 15 K in 20 kOe. Compared to the ZFC curve it shows an enhancement of the coercive field and a horizontal shift most likely due to EB. The ZFC curve shows a coercive field of 500 Oe, whereas the curve measured after FC displays an increased value of 1135 Oe. This is usually attributed to an additional uniaxial anisotropy contribution from short-range order correlations in the AF acting onto the FM [42-45].

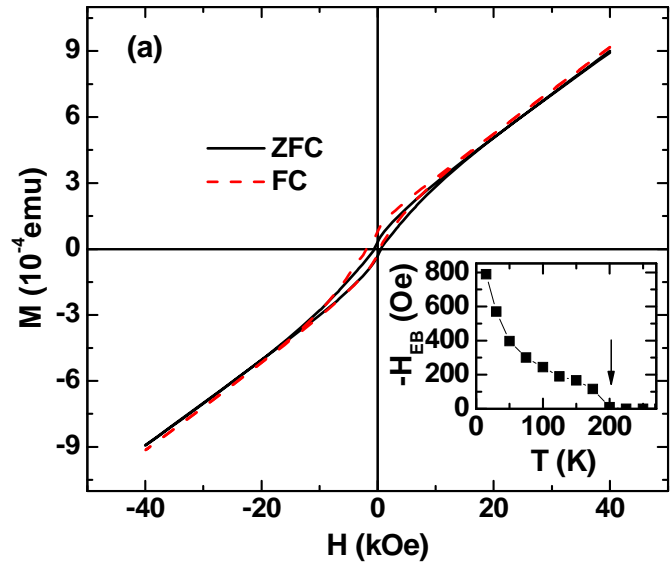

**FIG. 8.** $M$ vs. $H$ hysteresis curves at 15 K after ZFC (solid black line) and after FC (dashed red line) of a monolayer film of NPs dried at 80°C in air. The inset shows the EB field, $H_{EB}$, as extracted from hysteresis loops at various temperatures.

Furthermore, a vertical shift along the magnetization axes is observed. Vertical shifts of the hysteresis loops were observed for Ni/NiO [46], Co/CoO [47], and Fe/Fe-oxide core-shell NPs [48, 49], which is generally attributed to frozen spins in the shell. Simulation studies on core-shell NPs show that the microscopic origin of the vertical shift is due to the uncompensated pinned moments at the core-shell interface that facilitate the nucleation of non-uniform magnetic structures during the increasing field branch of the loops [37]. Additional evidence for the EB scenario can be obtained by measuring the temperature dependence of the exchange bias field $H_{EB}$ [36].

The inset of Fig. 8 shows the absolute value of $H_{EB}(T)$. The data were obtained by FC the sample in a magnetic field of 20 kOe from 330 K to each temperature. One observes that $H_{EB}$ decreases with increasing temperature and vanishes at $\approx$200 K. This matches well with the bulk Néel temperature of wüstite, $T_N$ = 198 K [19]. Therefore, structural (x-ray diffraction and TEM) together with magnetometric data consistently confirm a multi-phase system of wüstite-maghemite NPs.

Next we consider the monolayer film of NPs annealed at 170°C in air. Figure 9 (a) shows $M$ vs. $T$ curves after ZFC and after FC measured at 50 Oe. The ZFC curve shows a monotonic increase probably having a maximum at $T \approx$ 400 K (outside the accessible temperature range) as expected from the usual behavior of a SPM or SSG system. In contrast to the system dried at 80°C the curve does not display a sudden but a more smooth increase of the magnetization. The FC curve shows a behavior as expected for a SSG system, i.e. splitting from the ZFC curve near the maximum and a slight decrease with decreasing temperatures [31-32].

This behavior is consistent with the conclusions drawn from the x-ray diffraction pattern of the 170 °C annealed system, i.e. the NPs being either in a single phase of FiM $F_3O_4$ or $\gamma$-$Fe_2O_3$. Therefore the magnetic behavior can be explained by FiM NPs, which are coupled entirely by dipolar interaction. The magnetic property of each individual NP is SPM. However, due to strong enough inter-particle interactions the collective behavior resembles a SSG state [29-33]. This feature is not further discussed here but will be subject to a systematic study in a forthcoming paper [50].

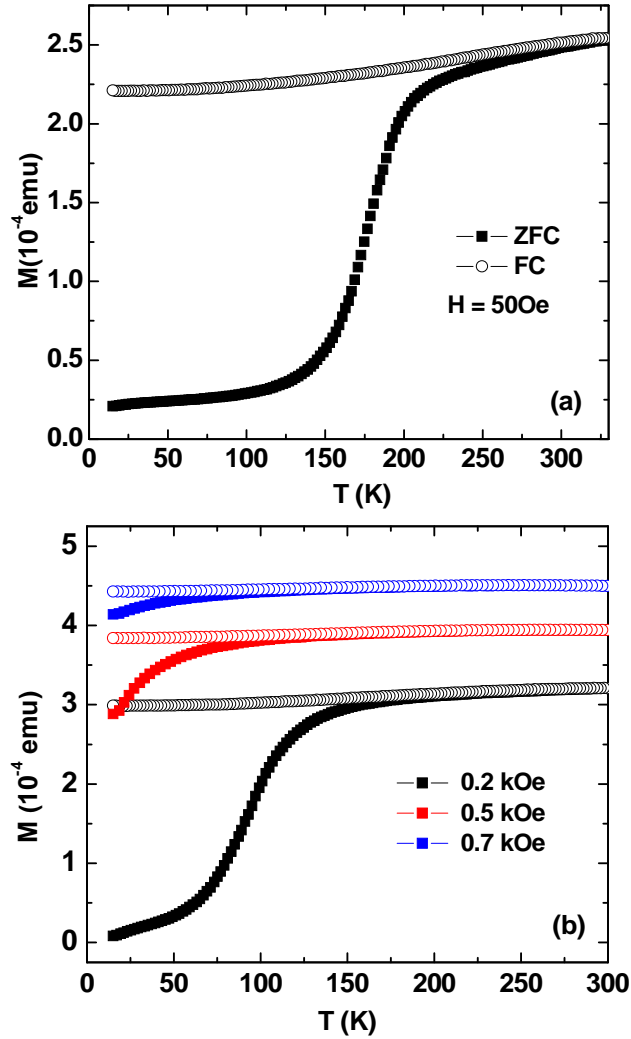

**FIG. 9.** *M* vs. *T* curves after ZFC (solid symbols) and after FC (open symbols) measured at (a) *H* = 50 Oe and (b) at various applied fields for a monolayer NPs annealed at 170°C in air. ZFC and FC measured at 0.5 and 0.7 kOe are shown with an offset of $2 \times 10^{-4}$ emu for better clarity.

Figure 9 (b) shows the magnetization vs. temperature after ZFC and FC measured at fields *H* = 0.2, 0.5 and 0.6 Oe. The peak in the ZFC curve shifts towards low temperatures as the field increases in contrast to the system discussed before. In larger fields the system becomes saturated (data not shown). Again, in the 170 °C annealed system we do not find indications for a Verwey transition at 120 K. We assume that also in these NPs at most only a small fraction of magnetite is present. Consequently, the composition of NPs after heat treatment at 170°C is mainly single-phase maghemite ($\gamma$-Fe$_2$O$_3$).

In Figure 10 (a) we show the hysteresis loops measured after ZFC at 15 K and 330 K. While at 15 K a FM-like open hysteresis loop is observed, at 330 K one finds a closed loop resembling that of a soft FM. This behavior is consistent with the assumption of pure FiM NPs, which are coupled by dipolar interaction.

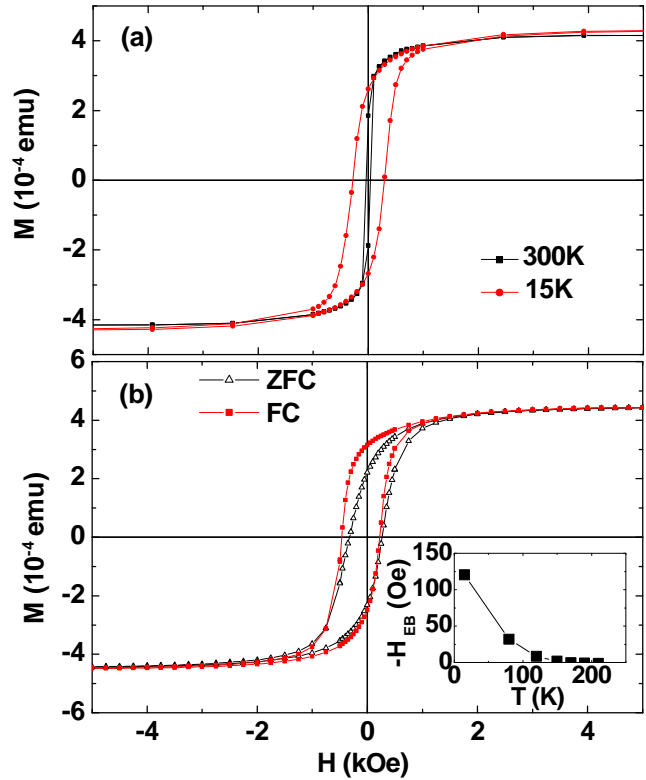

**FIG. 10.** *M* vs. *H* hysteresis curves at (a) 15 K (dashed red line) and 330 K (solid black line) after ZFC for a monolayer film of NPs annealed at 170°C in air and (b) at 15 K after ZFC (black open triangles) and FC (red solid squares). The inset shows the temperature dependence of the EB field.

However, the relatively sharp switching observed in the loop at 15 K does not correspond to a SSG scenario, where more rounded loops are expected [29-33]. Either a strong preferential orientation of anisotropy axes along the field axis is present or rather a superferromagnetic (SFM) [32, 33, 51-55] state is encountered. Again, since the collective NP behavior is not in the focus of the present work, the distinction between SSG or SFM case will not be analyzed further at this point [50].

Fig. 10 (b) compares the hysteresis loops at 15 K after ZFC or FC. Surprisingly, one observes a finite EB effect although no presence of AF wüstite was detected from structural data. The temperature dependence of $H_{EB}$ is plotted in the inset of Fig. 10 (b). The EB vanishes at $\approx$160 K and hence at a smaller temperature as compared to the case discussed above. It is very likely that this temperature corresponds to a finite-size scaled Néel temperature of wüstite and hence indicates the existence of very small size wüstite inside the otherwise FiM NPs that have escaped the x-ray diffraction analysis.

The third type of heat treatment was annealing the NPs at 230°C in vacuum. Fig. 11 (a) shows ZFC and FC curves measured at 50 Oe. Here we observe a clear step-like feature in both ZFC and FC curve at 110 K. These steps obviously mark the Verwey temperature indicative for the presence of a magnetite phase inside the NPs. The Verwey temperature is thereby reduced due to the finite-size effect as compared to the bulk value of 120 K.

Apart from the Verwey-transition the ZFC and FC curves look relatively similar to the second case of NPs that are annealed at 170°C in air, i.e. a smooth increase of the ZFC curve with a peak and a decrease in the FC curve upon cooling down. From the x-ray diffraction results the presence of AF wüstite is expected. However, no obvious effect of wüstite can be detected from the magnetization curves at low temperatures. The ZFC peak temperature ($\approx$300 K) is slightly smaller compared to the previous case signifying a smaller energy barrier for SPM fluctuations. This could be due to the fact that magnetite is the dominating phase inside the NPs with negligible magnetic influence of wüstite and hence less anisotropy from less EB effects and disorder. The decrease in the FC curve could again be an indication for dipolar coupling between NPs as in the case of annealing at 170°C in air.

On the one hand, the ZFC and FC curves measured at 0.2, 0.5 and 0.7 kOe (Fig. 11 (b)) show the expected decrease of the ZFC peak temperature with increasing field. For $H$ = 0.7 kOe the

ZFC peak seem to have shifted to ≈50 K. On the other hand, the Verwey transition remains for all fields at ≈110 K as expected for NPs consisting of magnetite.

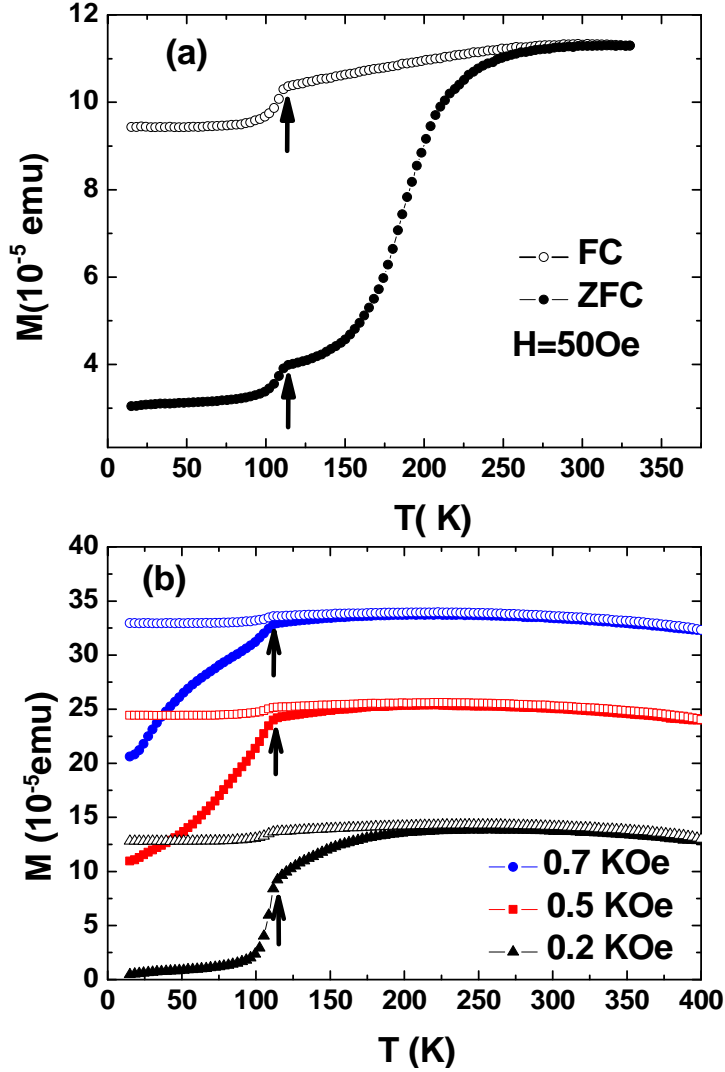

**FIG. 11.** $M$ vs. $T$ curves after ZFC (closed symbols) and after FC (open symbols)

measured at various applied fields for a monolayer film of NPs annealed at 230°C in vacuum. ZFC and FC measured at 50 Oe (a), 0.2, 0.5 and 0.7 kOe (b) are shown with an offset of $0.8 \times 10^{-4}$ and $1.5 \times 10^{-4}$ emu, respectively, for better clarity. The arrows indicate the Verwey transition temperature $T_v$ of magnetite at 110 K.

The magnetization hysteresis loops at 15 and 300 K are shown in Fig. 12 (a) . A very similar

behavior is found as for the system annealed at 170°C in air (see Fig. 9), i.e. a closed loop at

300 K similar to a soft-FM or to Stoner-Wohlfarth particles in the unblocked regime and a FM-type open hysteresis loop at 15 K again with a surprising squareness of the loop. This can also be interpreted in terms of a collective SFM state or Stoner-Wohlfarth behavior in the blocked regime with a preferential anisotropy axis.

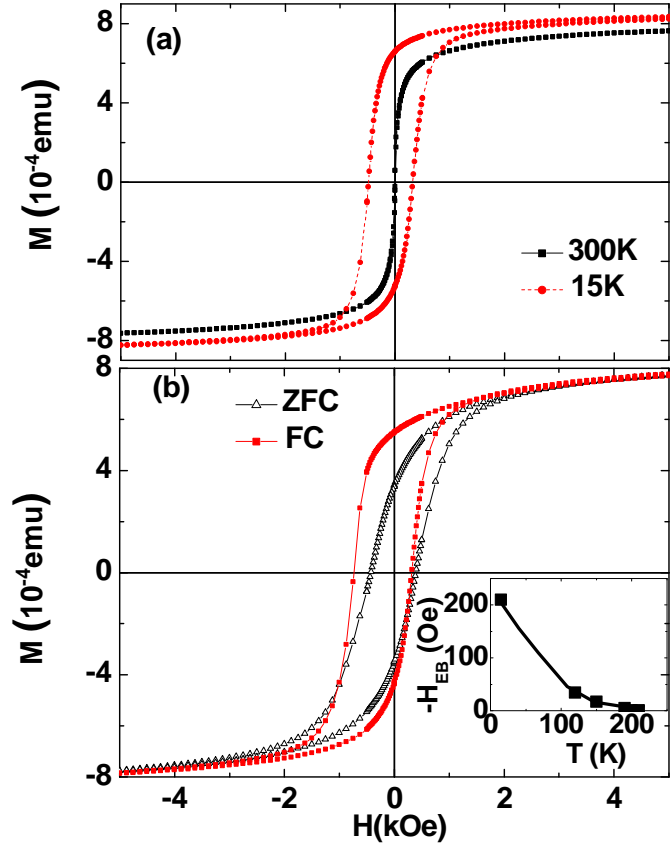

**FIG. 12. (a)** *M* vs. *H* hysteresis curves at 15 K (dashed red line) and 300 K (solid black line) after ZFC of a monolayer film of NPs annealed at 230 °C in vacuum and (b) *M* vs. *H* hysteresis curves at 15 K after ZFC (solid black line) and FC (dashed red line) in 500 Oe. The inset shows the temperature dependence of EB field. The solid line is a guide to the eyes.

Fig. 12(b) shows hysteresis loops at 15 K after ZFC and FC in 500 Oe. There is a clear difference between the ZFC and FC curves. In addition the FC curve displays an EB value of $H_{eb}$ = -210 Oe, while the ZFC curve shows zero EB as expected from usual EB systems [36, 37, 42-45]. The EB is again likely to be due to the interaction between AF wüstite and FiM magnetite inside each NP. This matches well with the x-ray studies that indicate the presence of wüstite.

The increased squareness of the loop after FC (first cycled loop) compared to the ZFC curve also corresponds to other EB systems [43-45]. In the case of FC a more or less well defined EB direction is imprinted into the wüstite phases during FC. This results in a square-like loop as also found in Fig. 12 (a) at 15 K.

However, in the ZFC case an isotropic distribution of EB directions is imprinted, which results in a rounded S-shaped loop. The loop after FC also shows a training effect (data not shown). The inset in Fig 12(b) shows the temperature dependence of $H_{EB}$ after FC in 500 Oe. The EB vanishes at ≈190 K, which is close to the Nèel temperature of wüstite (see above). This confirms that the EB effect results from the exchange interaction between wüstite and spinel phases inside each NP. Moreover, it signifies that the wüstite phase inside the NPs is larger in size than in the 170 °C annealing case but larger compared to the 80°C dried case.

Further information about the magnetic behavior after all three annealing processes can be drawn from so-called thermoremanent (TRM) and iso-thermoremanent (IRM) magnetization data [35].

In the TRM protocol the system is cooled down from high temperatures in a certain applied field, $H$. At the target temperature (e.g. below the blocking, spin glass or a transition temperature, respectively) the field is switched off and the remanent magnetization recorded. This value is the TRM value corresponding to the field $H$ at this specified temperature. In the IRM protocol, however, the system is cooled down in zero field down to the target temperature, then a field $H$ is momentarily applied, subsequently switched off and eventually the remanent magnetization is recorded. This value is the IRM corresponding to the field $H$. Consequently, in non-ergodic systems, TRM and IRM values should be non-zero and often differ from each other since two different histories are passed. It has been shown that TRM/IRM plots vs. applied field $H$ can be used as 'fingerprints' for distinguishing between different classes of magnetic properties [35, 56].

The TRM/IRM plot at 15 K of the NPs dried at 80°C in air is shown in Fig. 13(a). The TRM increases monotonically for small fields as expected for SPM or SSG type of systems [35, 56]. At ≈10 kOe a broad peak is found and for larger fields there is a weak decrease.

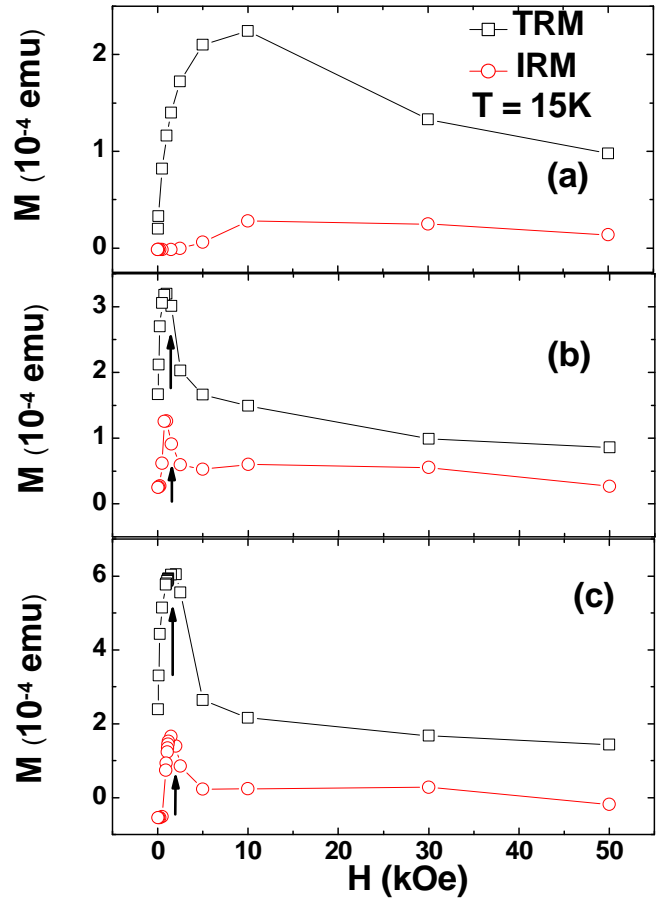

**FIG 13.** TRM and IRM vs. applied field measured at 15 K of a monolayer film of NPs annealed at (a) 80°C in air (b) 170°C in air and (c) 230 °C in vacuum, respectively.

This behavior has been reported previously for spin glass (SG) systems [57, 58]. Consequently it might be attributed here to SSG type of behavior of dipolarly coupled NPs. However, the IRM stays at relatively small values, which is atypical for SG and hence SSG systems. This might be due to the presence of AF wüstite. An ideal AF system shows zero IRM for all fields [56] and therefore the overall IRM value is reduced.

Fig. 13 (b) and (c) shows the TRM/IRM curves at 15 K of the samples annealed in 170°C in air and 230°C in vacuum, respectively. For small field values the TRM and IRM curves follow again the expectation for a SPM or SSG system [56], i.e. both are monotonically increasing while IRM < TRM. However, at intermediate fields one finds in both systems a surprisingly sharp maximum in TRM and IRM with the IRM peak slightly below the TRM peak. At 15 K the TRM has a peak at a field of ≈1000 Oe (170°C in air) and at ≈1500 Oe (230°C in vacuum), respectively. For larger fields both TRM and IRM decrease drastically and then saturate for fields $H \rightarrow 50$ kOe. Such a sharp peak in TRM and IRM has not been discussed in the literature yet to the best of our knowledge.

This behavior implies that for fields larger than the peak value ($H_{peak,TRM}$ and $H_{peak,IRM}$, respectively) the remanent magnetization decreases, despite the fact that the applied field is increased. This can be explained by two field regimes, where the magnetization relaxation process occurs with different relaxation times. Below the peak value the behavior is regular with increasing remanence with increasing fields. Above the peak, the relaxation obviously proceeds much faster than in the first regime. This can be understood considering the $H$-$T$ phase diagram for wüstite. Unfortunately, no experimental phase diagram for wüstite is known to our best knowledge. Since $Fe_xO$ is an AF, the $H$-$T$ diagram will likely resemble the schematic depicted in the inset of Fig. 14. A phase line (usually a phase line of second order and one of first order being separated by a multicritical point [59]) separates the AF phase from the paramagnetic (PM) phase.

For FC in relatively small fields one enters the AF phase. At low temperatures (in our case 15 K) the field is switched off and then the TRM value is recorded. The system stays in the AF phase when the field is switched off. Therefore, the magnetization relaxation is governed by a slow relaxation to the AF ground state via AF domain wall dynamics [60] with some relaxation time $\tau_2$. However, for FC in large fields one stays in the PM phase during cooling. When the field is switched off, the systems crosses the AF phase boundary, $H_{crit}(T)$, enters the AF phase and relaxes toward the AF ground state. This relaxation path obviously occurs with a smaller relaxation time, $\tau_1 < \tau_2$, than the previous one.

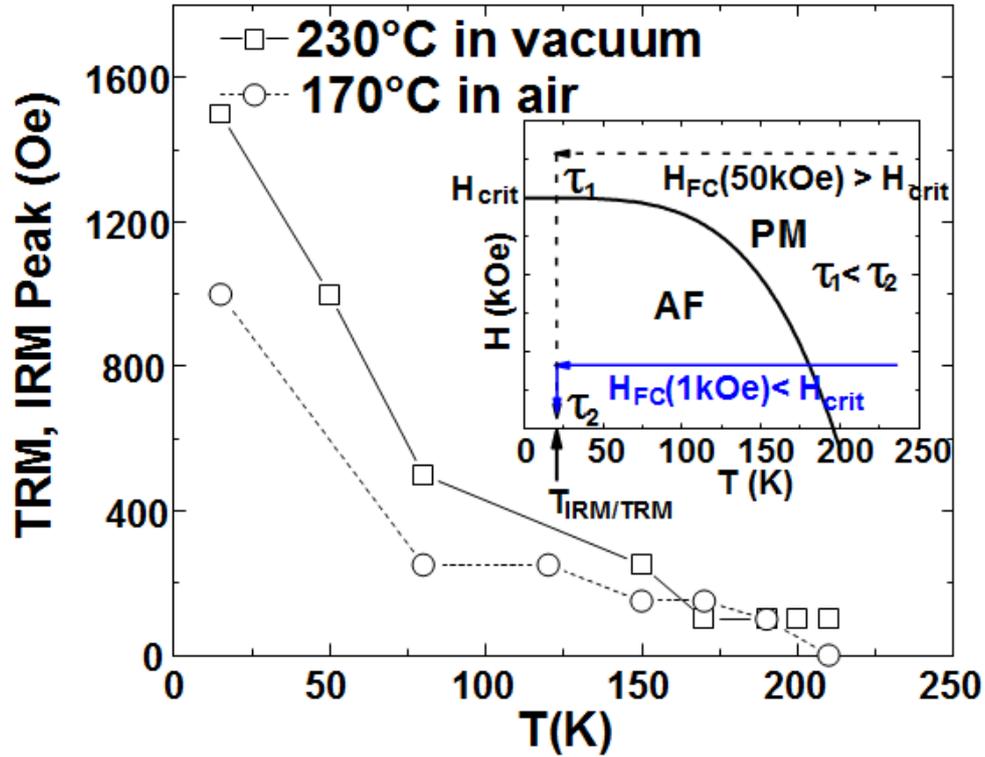

**FIG. 14**. Peak field value from the TRM plot (see Fig. 13) as function of temperature of a monolayer of NPs annealed at 170°C in air and 230°C in vacuum, respectively. The inset shows a qualitative phase diagram of wüstite. The black solid line marks the phase boundary $H_{crit}(T)$. The arrows depict two different FC approaches as examples, i.e. FC with 1 kOe and thus entering the AF phase and FC in 50 kOe, which stays outside the AF phase. The magnetization relaxation at 15 K then occurs with two different relaxation times, $\tau_1$ and $\tau_2$, respectively, as described in the text.

Consequently, the peak in TRM and IRM marks the AF phase boundary. This scenario is supported by measuring the peak field of the TRM curve as function of temperature. If the peak coincides with $H_{crit}(T)$, then it should decrease with increasing temperature and eventually become zero near the Néel temperature. This behavior is confirmed in Fig. 14. For the system annealed in 230°C in vacuum one clearly observes a monotonic decrease from ≈1500 Oe down to 100 Oe at ≈170 K, which is near to the finite-size Néel temperature of wüstite extracted above, i.e. ≈190 K. For the case of annealing at 170°C in air a less clear trend is observable. Nevertheless, also a monotonic decrease from ≈1000 Oe down to zero at 200 K is found. However, the curve stays systematically below the one of the 230°C annealed case. One can roughly estimate from the shape of the curve a characteristic temperature of ≈160K, which would match with the stronger finite-size scaled Néel temperature of wüstite as inferred above.

Comparing and summarizing all data on the three annealing studies we conclude that in all three cases a mixed-phase systems is found., i.e. 80°C dried in air (wüstite-maghemite), 170°C in air (mainly maghemite with a residual wüstite component), and 230°C in vacuum (wüstite-magnetite). Moreover, one can infer that in the first case the wüstite phase is much larger in size compared to the second and third one. This leads to a larger (finite size scaled) $H_{crit}(T)$ in the $H$-$T$ phase diagram of wüstite being outside of the accessible field range. Therefore, no sharp peak is found in the TRM plot in this case. In the second and third annealing case, the $H_{crit}(T)$ is at intermediate fields and thus two different relaxation paths are possible. Consequently, a sharp peak can be detected in the TRM.

## IV. CONCLUSIONS

In summary, we have investigated self-assembled iron oxide NPs with a diameter of 20 nm and a size distribution of 7%. After spin-coating on Si substrates mono- or multilayer films of self-organized NPs were established. The crystallographic structure of the NPs and thus the magnetic behavior strongly depends on the thermal treatment of the NP films. Drying in 80°C in air is equivalent to the as-prepared state and yields a multi-phase system of $Fe_xO/\gamma$-$Fe_2O_3$ NPs. The magnetic behavior is characterized by both SPM blocking and EB between the AF $Fe_xO$ and FiM $\gamma$-$Fe_2O_3$. TEM studies suggest a polycrystalline multi-phase instead of a core-shell structure. Annealing in 170°C in air yields a predominantly γ-$Fe_2O_3$ structure of NPs. The magnetic behavior is governed by dipolarly coupled FiM NPs. From magnetometry and in particular TRM/IRM measurements a small wüstite component is detected here as well. The third heat treatment, i.e. 230°C in vacuum produces multi-phase NPs of $Fe_xO/Fe_3O_4$. The presence of magnetite is confirmed by the Verwey transition. The system is also characterized by SPM blocking and EB between the AF $Fe_xO$ and FiM $Fe_3O_4$.

ACKNOWLEDGEMENTS

M.J.B. acknowledges support from the International Max-Planck Research School "SurMat" and D.M. from the NRW Graduate School: "Research with synchrotron radiation in the nano- and biosciences", Dortmund.

# Supplementary data

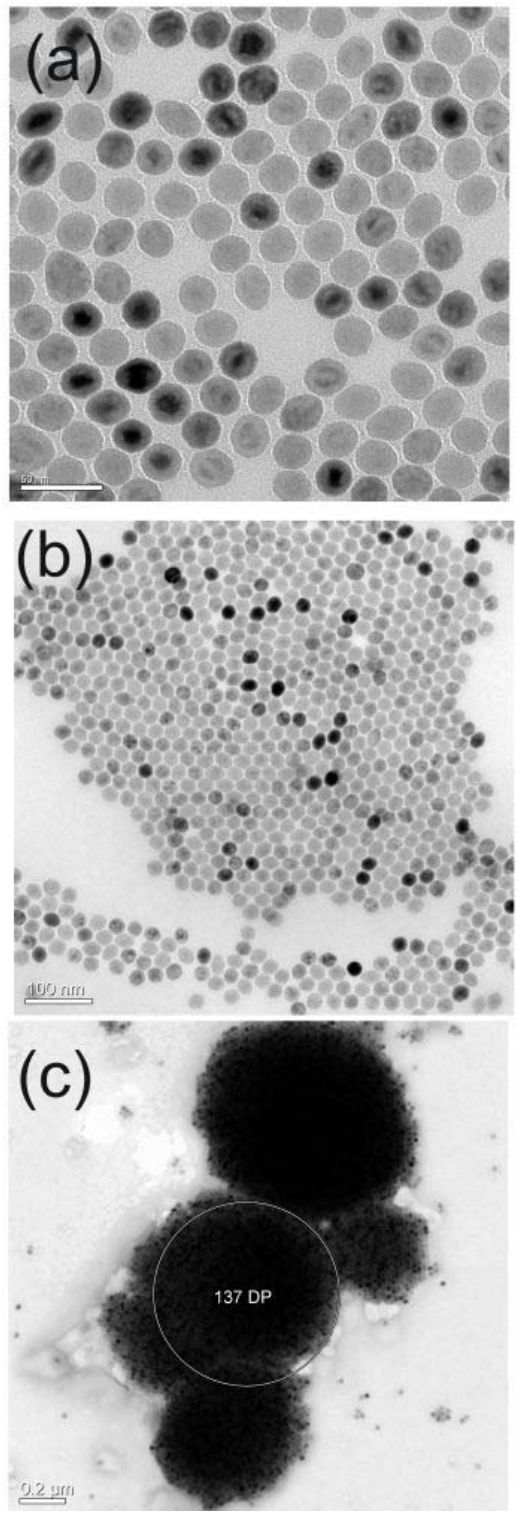

Fig S1: TEM image of the NP area illuminated for SAED patterns for NPs annealed at (a) 80°C in air (b) 170°C in air (c) 230°C in vacuum.